\documentclass[twocolumn,secnumarabic,amssymb, nobibnotes, aps, prl]{revtex4-1}

\usepackage[applemac]{inputenc}

\usepackage[T1]{fontenc}
\usepackage[english]{babel}
\usepackage{layout}
\usepackage[top=2.5cm, bottom=2.cm, left=2.cm, right=2.cm]{geometry}
\usepackage{float}
\restylefloat{table}
\usepackage{amsthm}
\usepackage{amsmath}
\usepackage{amssymb}
\usepackage{mathrsfs}
\usepackage{stmaryrd}
\usepackage[usenames]{xcolor}
\usepackage[backref=false]{hyperref}
\hypersetup{
colorlinks=true,
citecolor=red,
linkcolor=darkblue
}
\definecolor{darkblue}{rgb}{0,0,.8}
\definecolor{red}{rgb}{1,0,0}
\usepackage{graphicx}
\usepackage{dsfont}

\newcommand{\nc}{\newcommand}
\nc{\be}{\begin{equation}}
\nc{\ee}{\end{equation}}
\nc{\bea}{\begin{eqnarray}}
\nc{\eea}{\end{eqnarray}}
\nc{\rc}[1]{\textcolor{red}{#1}}
\nc{\bra}[1]{\langle{#1}|}
\nc{\ket}[1]{|{#1}\rangle}
\nc{\hbdy}{h_{\rm bdy}}
\nc{\hbdyvec}{{\bf h}_{\rm bdy}}

\DeclareMathOperator{\Tr}{Tr}
\DeclareMathOperator{\STr}{STr}
\DeclareMathOperator{\MTr}{MTr}
\DeclareMathOperator{\Arg}{Arg}

\usepackage{tikz}
\usetikzlibrary{calc}
\usetikzlibrary{chains}
\usetikzlibrary{scopes}
\usetikzlibrary{shapes.geometric}
\usetikzlibrary{trees}
\usetikzlibrary{topaths}
\usetikzlibrary{positioning}
\usetikzlibrary{patterns,decorations.pathreplacing}

\newcommand{\gsNtwoInline}{
\begin{tikzpicture}[scale=0.25,baseline=1pt]
 \draw[thick] (1,1.) arc (0:-180:0.5);
\end{tikzpicture}
}

\newcommand{\TLGenInline}{
\begin{tikzpicture}[scale=0.25,baseline=1pt]
 \draw[thick] (1,1.) arc (0:-180:0.5);
 \draw[thick] (1,-0.2) arc (0:180:0.5);
 \end{tikzpicture}
}

\newcommand{ \IdentityTwoSites}{
\begin{tikzpicture}[baseline=4pt,scale=0.3]
 \draw[thick] (0,0) -- (0,1.5);
 \draw[thick] (1,0) -- (1,1.5);
  \end{tikzpicture}
 }

\newcommand{ \IdentityTwoSitesInline}{
\begin{tikzpicture}[baseline=1pt,scale=0.25]
 \draw[thick] (0,0) -- (0,1);
 \draw[thick] (1,0) -- (1,1);
  \end{tikzpicture}
 }

\newcommand{ \DensityMatTwoSites}{
\begin{tikzpicture}[baseline=4pt,scale=0.3]
 \draw[thick] (0,1.5) arc (0:-180:0.5);
 \draw[thick] (0,0) arc (0:180:0.5);
 \end{tikzpicture}
 }
 
\newcommand{ \ReducedDensityMatOneSite}{\quad
\begin{tikzpicture}[baseline=3pt,scale=0.5]
 \draw[thick] (0,-0.1) -- (0,.9);
 \end{tikzpicture}\quad
 }

\newcommand{ \ReducedDensityMatOneSiteInline}{
\begin{tikzpicture}[scale=0.25,baseline=1pt]
 \draw[thick] (0,-0.1) -- (0,.9);
 \end{tikzpicture}
 }
 
\newcommand{ \linkStateA}{\quad
\begin{tikzpicture}[baseline=-6pt,scale=0.3]
 \draw[thick] (0,0) arc (-180:0:0.5);
 \draw[thick] (2,0) arc (-180:0:0.5);
 \end{tikzpicture}\quad
 }
 
 \newcommand{ \linkStateB}{\quad
\begin{tikzpicture}[baseline=-6pt,scale=0.3]
 \draw[thick] (1,0) arc (-180:0:0.5);
 \draw[thick] (0,0) arc (-180:0:1.5 and 1.0);
 \end{tikzpicture}\quad
 }
 
  \newcommand{ \wordA}{\quad
\begin{tikzpicture}[baseline=10pt,scale=0.3]
 \draw[thick] (0,1.5) arc (-180:0:0.5);
 \draw[thick] (2,1.5) arc (-180:0:0.5);
  \draw[thick] (0,0) arc (180:0:0.5);
 \draw[thick] (2,0) arc (180:0:0.5);
  \end{tikzpicture}\quad
  }
  
\newcommand{\wordB}{\quad
\begin{tikzpicture}[baseline=10pt,scale=0.3]
 \draw[thick] (1,0) arc (180:0:0.5);
 \draw[thick] (0,0) arc (180:0:1.5 and 0.8);
 \draw[thick] (0,1.5) arc (-180:0:0.5);
 \draw[thick] (2,1.5) arc (-180:0:0.5);
\end{tikzpicture}\quad
 }
 
 \newcommand{\wordC}{\quad
\begin{tikzpicture}[baseline=10pt,scale=0.3]
 \draw[thick] (1,1.5) arc (-180:0:0.5);
 \draw[thick] (0,1.5) arc (-180:0:1.5 and 0.8);
  \draw[thick] (0,0) arc (180:0:0.5);
 \draw[thick] (2,0) arc (180:0:0.5);
\end{tikzpicture}\quad
 }
 
\newcommand{\wordD}{\quad
\begin{tikzpicture}[baseline=10pt,scale=0.3]
 \draw[thick] (1,1.5) arc (-180:0:0.5 and 0.3);
 \draw[thick] (0,1.5) arc (-180:0:1.5 and 0.6);
 \draw[thick] (1,0) arc (180:0:0.5 and 0.3);
 \draw[thick] (0,0) arc (180:0:1.5 and 0.6);
\end{tikzpicture}\quad
 } 
 
\begin{document}
\title{Entanglement in non-unitary quantum critical spin chains}

\author{Romain Couvreur$^{1,3}$, Jesper Lykke Jacobsen$^{1,2,3}$ and  Hubert Saleur$^{3,4}$}

\affiliation{${}^1$Laboratoire de Physique Th\'eorique, \'Ecole Normale Sup\'erieure -- PSL Research University, 24 rue Lhomond, F-75231 Paris Cedex 05, France}
\affiliation{${}^2$Sorbonne Universit\'es, UPMC Universit\'e Paris 6, CNRS UMR 8549, F-75005 Paris, France} 
\affiliation{${}^3$Institut de Physique Th\'eorique, CEA Saclay, 91191 Gif Sur Yvette, France}
\affiliation{${}^4$Department of Physics and Astronomy, University of Southern California, Los Angeles, CA 90089-0484}

\date{\today}

\begin{abstract}
Entanglement entropy has proven invaluable to our understanding of quantum criticality. It is natural to try to extend the concept to  ``non-unitary quantum mechanics'', which has seen growing interest from areas as diverse as open quantum systems, non-interacting electronic disordered systems, or non-unitary conformal field theory (CFT). We propose and investigate such an extension here, by focussing on the case of one-dimensional quantum group symmetric or supergroup symmetric spin chains. We show that the consideration of left and right eigenstates combined with appropriate definitions of the trace leads to a natural definition of R\'enyi entropies in a large variety of models. We interpret this definition geometrically in terms of related loop models and calculate the corresponding scaling in the conformal case. This allows us to distinguish the role of the central charge and effective central charge in rational minimal models of CFT, and to define an effective central charge in other, less well understood cases. The example of the $sl(2|1)$ alternating spin chain for percolation is discussed in detail.
\end{abstract}


\maketitle

The concept of entanglement entropy has profoundly affected our understanding of quantum systems, especially in the vicinity of critical points \cite{JPHYSAspecial}.
A growing interest in {\em non-unitary quantum mechanics} (with non-hermitian ``Hamiltonians'') stems from open quantum systems, where
the reservoir coupling can be represented by hermiticity-breaking boundary terms \cite{Prozen}. Another motivation comes from disordered
non-interacting electronic systems in $2+1$ dimensions (D) where phase transitions, such as the plateau transition in the integer quantum
Hall effect (IQHE), can be investigated---using a supersymmetric formalism and dimensional reduction---via 1D non-hermitian
quantum spin chains with supergroup symmetry (SUSY) \cite{Zirnbauer}. SUSY spin chains and quantum field theories with target space
SUSY also appear in the AdS/CFT correspondence \cite{Beisert,Volkerreview} and in critical  geometrical systems such as polymers or
percolation \cite{Parisi}. Quantum mechanics with non-hermitian but PT-symmetric ``Hamiltonians'' also gains increased interest \cite{Bender}.

Can entanglement entropy be meaningfully extended beyond ordinary quantum mechanics? We focus in
this Letter on critical 1D spin chains and the associated 2D critical statistical systems and CFTs. This is the area where our understanding of
the ordinary case is the deepest, and the one with most immediate applications.  

For ordinary critical quantum chains (gapless, with linear dispersion relation), the best known result concerns the entanglement entropy (EE) of a subsystem
$A$ of length $L$ with the (infinite) rest $B$ at temperature $T=0$. Let $\rho_A=\hbox{Tr}_B \rho$ denote the reduced density operator, where
$|0\rangle$ is the normalized ground state and $\rho=|0\rangle\langle 0|$. The (von Neumann) EE then reads $S_A=-\hbox{Tr}_A \rho_A\ln\rho_A$.
One has $S\approx {c\over 3}\ln (L/a)$ for $L \gg a$, where $a$ is a lattice cutoff and $c$ the central charge of the associated CFT.
For the XXZ chain, $c=1$. 

Statistical mechanics is ripe with non-hermitian critical spin chains:\ the Ising chain in an imaginary magnetic field (whose critical point is described
by the Yang-Lee singularity), the alternating $sl(2|1)$ chain describing percolation hulls \cite{ReadSaleur01}, or the alternating $gl(2|2)$ chain
describing the IQHE plateau transition \cite{Zirnbauer}. The Ising chain is conceptually the simplest, as it corresponds to a {\sl rational} non-unitary CFT.
In this case, abstract arguments \cite{Doyon1,Doyon2} suggest replacing the unitary result by 
\begin{equation}
S_A\approx {c_{\rm eff} \over 3}\ln (L/a) \,, \label{ceffres}
\end{equation}
where $c_{\rm eff}$ is the {\sl effective} central charge. For instance, for the Yang-Lee singularity, $c=-{22\over5}$ but $c_{\rm eff}={2\over 5}$;
in this case (\ref{ceffres}) was checked numerically \cite{Doyon1}. It was also checked analytically for integrable realizations
of the non-unitary minimal CFT. 
%
%
The superficial similarity with the result $s\approx {\pi c_{\rm eff}\over 3}T$ for the thermal entropy per unit length of the infinite chain at $T \ll 1$
suggests that (\ref{ceffres}) is a simple extension of the scaling of the ground-state energy in non-unitary CFT \cite{ISZ}. But the situation is more
subtle, as can be seen from the fact that the leading behavior of the EE is independent of the (low-energy) eigenstate in which it is computed \cite{Sierra}. 

There are two crucial conditions in the derivation of  (\ref{ceffres}): the left and right ground states $|0_{\rm L}\rangle, |0_{\rm R}\rangle$ must be identical,
and  the full operator content of the theory must be  known. These conditions hold for minimal, rational CFT, but in the vast majority of systems
the operator content depends on the boundary conditions (so it is unclear what  $c_{\rm eff}$ is), and $|0_{\rm L}\rangle\neq |0_{\rm R}\rangle$, begging
the question of how exactly $\rho$, $\rho_A$ and $S_A$ are defined. 

In this Letter we explore this vast subject by concentrating on non-Hermitian models with SUSY or quantum group (QG) symmetry.
We extend the general framework of Coulomb gas and loop model representations to EE calculations. We derive
(\ref{ceffres}) for minimal non-unitary models, and define modified EE involving the true $c$ even in non-unitary cases. We finally
introduce a natural, non-trivial EE in SUSY cases, even when the partition function $Z=1$. 

\paragraph{EE and QG symmetry.} We first discuss the critical $U_qsl(2)$ QG symmetric XXZ spin chain \cite{PS}.
Let $\sigma_i^{x,y,z}$ be Pauli matrices acting on space $i$ and define the nearest neighbor interaction
\begin{equation}
e_i=-\tfrac{1}{2} \left[ \sigma_i^x\sigma_{i+1}^x+\sigma_i^y\sigma_{i+1}^y
+ \tfrac{q+q^{-1}}{2}(\sigma_i^z\sigma_{i+1}^z-1) + h_i \right] \nonumber
\end{equation}
with $q\in \mathbb{C}$, $|q|=1$. The Hamiltonian $H=-\sum_{i=1}^{M-1} e_i$ with $h_i=0$ describes the ordinary critical XXZ chain on $M$ sites, but we add the
hermiticity-breaking boundary term $h_i =  \tfrac{q-q^{-1}}{2}(\sigma_i^z-\sigma_{i+1}^z)$ to ensure commutation with the $U_qsl(2)$ QG
(whose generators are given in the supplemental material (SM)).

Consider first 2 sites, that is $H=-e_1$. $H$ is not hermitian; its eigenvalues are real \cite{Staubin} but its left and right eigenstates differ.
We restrict $\Arg q \in [0,\pi/2]$, so the lowest energy is $E^{(0)}=-(q+q^{-1})$ (the other eigenenergy is $E^{(1)}=0$).
The right ground state, defined as $H|0\rangle=E^{(0)}|0\rangle$, is
$|0\rangle= \tfrac{1}{\sqrt{2}}(q^{-1/2}|\uparrow \downarrow \rangle-q^{1/2}|\downarrow \uparrow \rangle)$.
We use the (standard) convention that complex numbers are conjugated when calculating the bra associated with a given ket;
therefore $\langle0|0\rangle=1$. The density matrix
\begin{equation}
\rho=|0\rangle\langle0|={1\over 2}\left(
\begin{smallmatrix}
0 & 0 & 0 & 0 \\
0 &1 &-q^{-1} & 0 \\
0 &-q &1 & 0 \\
0 & 0 & 0 & 0
\end{smallmatrix}\right)
\end{equation}
(in the basis $\uparrow\uparrow,\uparrow\downarrow,\downarrow\uparrow,\downarrow\downarrow$)
is normalized, $\Tr \rho = 1$. Taking subsystem A (B) as the left (right) spin, the reduced density operator is
$\rho_A= \tfrac{1}{2} \left(\begin{smallmatrix} 1 & 0 \\ 0 & 1 \end{smallmatrix} \right)$,
and therefore 
\begin{equation}
S_{A}=\ln 2 \,. \label{Svn}
\end{equation}
This coincides with the well-known result for the  $sl(2)$ symmetric (hermitian) XXX chain ($q=1$).
But since $H$ is non-hermitian, it is more correct to work with left and right eigenstates defined by
$H|E_{\rm R}\rangle=E|E_{\rm R}\rangle$ and $\langle E_{\rm L}|H=E\langle E_{\rm L}|$ 
(or $H^\dagger |E_{\rm L}\rangle=E|E_{\rm L}\rangle$,
since $E \in \mathbb{R}$). Restricting to the sector $S^z=0$ we have 
\begin{eqnarray}
|0_{\rm R}\rangle &=& \tfrac{1}{\sqrt{q+q^{-1}}}\left(q^{-1/2}|\uparrow\downarrow \rangle-q^{1/2}|\downarrow\uparrow \rangle\right) \label{0Rstate} \\
|1_{\rm R}\rangle &=& \tfrac{1}{\sqrt{q+q^{-1}}}\left(q^{1/2}|\uparrow\downarrow \rangle+q^{-1/2}|\downarrow\uparrow \rangle\right) \label{1Rstate}
\end{eqnarray}
where $|0_{\rm R}\rangle$, $|1_{\rm R}\rangle$ denote the right eigenstates with energies $E^{(0)}, E^{(1)}$.
The left eigenstates $|0_{\rm L}\rangle$, $|1_{\rm L}\rangle$ are obtained from (\ref{0Rstate})--(\ref{1Rstate}) by $q \to q^{-1}$.
%
%
Normalizations are such that $\langle i_{\rm L}|i_{\rm R}\rangle=1$, and $\langle i_{\rm L}|j_{\rm R}\rangle=0$ for $i\neq j$.
Since $\langle 0_{\rm R}|1_{\rm R}\rangle\neq0$ we need both L and R eigenstates to build a projector onto the ground state.
We thus {\em define}
\begin{equation}
\tilde{\rho}\equiv |0_{\rm R}\rangle\langle 0_{\rm L}|= \frac{1}{q+q^{-1}}\left(\begin{smallmatrix}
0 & 0 & 0 & 0 \\
0 &q^{-1} & -1 & 0 \\
0 & -1 & q & 0 \\
0 & 0 & 0 & 0
\end{smallmatrix}\right) \,,
\end{equation}
and $\tilde{\rho}_A= \Tr_B \left(q^{-2\sigma_{B}^z}\tilde{\rho}\right)= \tfrac{1}{q+q^{-1}}\left(\begin{smallmatrix} 1 & 0 \\ 0 & 1 \end{smallmatrix}\right)$.
We justify the use of a modified trace shortly with both geometrical and QG considerations.
Observe that $\tilde{\rho}_A$ is normalized for the modified trace (note the opposite power of $q$):
$\Tr_A \left(q^{2\sigma_A^z} \tilde{\rho}_A \right)=1$. We now define the EE as 
\begin{equation}
\tilde{S}_{A}=-\Tr \left(q^{2\sigma_A^z}\tilde{\rho}_A\ln\tilde{\rho}_A\right)=\ln(q+q^{-1}) \,. \label{Svnmod}
\end{equation}
The result (\ref{Svnmod}) is more appealing that (\ref{Svn}): it depends on $q$ through the combination $q+q^{-1}$
which is the quantum dimension of the spin $1/2$ representation of $U_qsl(2)$.
Note that (\ref{Svnmod}) satisfies $\tilde{S}_A=\tilde{S}_B$ (see SM).

%
%

\paragraph{Entanglement and loops.} Eq.~(\ref{Svnmod}) admits an alternative interpretation in terms of loop models.
Since $e_i$ obey the Temperley-Lieb (TL) relations,
\begin{eqnarray}
e_i^2 &=& (q+q^{-1})e_i \,, \nonumber \\
e_i e_{i \pm 1} e_i &=& e_i \,, \nonumber\\
 \left[e_i,e_j\right] &=& 0 \mbox{ for } |i-j|>1 \,, \label{TLrels}
\end{eqnarray}
their action can be represented in terms of diagrams:  $e_i = \TLGenInline$ contracts neighboring lines, and multiplication means stacking diagrams
vertically, giving weight $n \equiv q+q^{-1}$ to each closed loop.
The ground state of $H=-e_1$ is
$\ket{0_\ell} = \tfrac{1}{\sqrt{n}} \gsNtwoInline$ ($\ell$ stands for loop).
We check graphically that $H \ket{0_\ell}=-n\ket{0_\ell}$.
With the scalar product ordinarily used in loop models (see SM), $\ket{0_\ell}$ is correctly normalized.
The density matrix is 
$\rho_\ell = \tfrac{1}{n} \ket{0_\ell}\bra{0_\ell} = \tfrac{1}{n} \TLGenInline$.
The partial trace $\rho_{A,\ell} = \Tr_B \rho_\ell$ glues corresponding sites on top and bottom throughout $B$ (here site $2$).
The resulting reduced density matrix acts only on $A$ (site $1$):
$\rho_{A,\ell} = \tfrac{1}{n} \ \ReducedDensityMatOneSiteInline \ $.
The gluing of $A$ creates a loop of weight $n$, so
$S_{A,\ell} = -\Tr (\rho_{A,\ell}\log\rho_{A,\ell}) = - n \times \frac{1}{n}\log\frac{1}{n} = \log n$.
The agreement with (\ref{Svnmod}) is of course no accident. Indeed, for any spin-$1/2$ Hamiltonian expressed in the
TL algebra (and thus commuting with $U_qsl(2)$), the EE---and in fact, the $N$-replica R\'enyi (see below)
entropies---obtained with the modified traces and with the loop construction coincide. We shall call these
{\sl QG entropies}, and denote them $\tilde{S}$.

\paragraph{Coulomb gas calculation of the EE.} For the critical QG invariant XXZ chain
with $H=-\sum e_i$, the EE $\tilde{S}$ scales as expected in CFT, but with the {\sl true} central charge 
$c=1-{6\over x(x+1)}$ (instead of $c_{\rm eff}=1$), where we parametrized $q={\rm e}^{i\pi/(x+1)}$. 
The simplest argument for this claim is field theoretical. We follow \cite{CardyCalabrese},
where the R\'enyi EE, $S^{(N)}\equiv {1\over 1-N}\ln \Tr \rho^N$, is computed from $N$ copies
of the theory on a Riemann surface with two branch points a distance $L$ apart.  As the density operator is obtained by
imaginary time evolution, we must  project, in the case of non-unitary CFT, onto $|0_{\rm R}\rangle$ in the ``past''
and on $|0_{\rm L}\rangle$ in the ``future'', to obtain $\tilde{\rho}=|0_{\rm R}\rangle\langle 0_{\rm L}|$. 

%

We calculate the QG R\'enyi EE using the loop model. The  geometry of \cite{CardyCalabrese} leads to a simple generalization
of well-known partition function calculations \cite{DFSZ}:\ an ensemble of dense loops now lives on $N$ sheets (with a cut of length $L$),
and {\sl each loop} has weight $n$. Let $Z^{(N)}$ denote the partition function. Crucially, there are now {\sl two types} of loops:
those which do not intersect the cut close after winding an angle $2\pi$, but those which do close after winding $2N\pi$.
To obtain the R\'enyi EE, we must find the dependence of $Z^{(N)}$ on $L$.

To this end we use the Coulomb gas (CG) mapping \cite{Nienhuis,JesperReview}.
The TL chain is associated with a model of oriented loops on the square lattice. Assign a phase ${\rm e}^{\pm i e_0/4}$ to each left (right) turn. 
In the plane, the number of left minus the number of right turns is $\Delta N_\pm = \pm 4$, so the weight $n=2\cos e_0$ results from summing over orientations.
The oriented loops then provide a vertex model, hence a solid-on-solid model on the dual lattice. Dual height variables are defined by
induction, with the (standard) convention that the heights across an oriented loop edge differ by $\pi$. In CG theory,
the large-distance dynamics of the heights is described by a Gaussian field $\phi$ with action
$A[\phi]={g\over 4\pi}\int {\rm d}^2x  \left[(\partial_x\phi)^2+(\partial_y\phi)^2\right]$
and coupling $g=1-e_0={x\over x+1}$. 

With $N$ replicas, we get in this way $N$ bosonic fields $\phi_1,\ldots,\phi_N$. The crux of the matter is the cut: a loop winding $N$ times
around one of its ends should still have weight $n$, whilst, since $\Delta N_\pm = \pm 4N$ on the Riemann surface, it gets instead $n'=2\cos N\pi e_0$.
We repair this by placing {\em electric charges} at the two ends (labelled $l,r$) of the cut, $e_l=e-e_0$ and $e_r=-e-e_0$, where
$e$ will be determined shortly. More precisely,  we must  insert the vertex operators $\exp[ie_{l,r} (\phi_1+\ldots+\phi_N)(z_{l,r},\bar{z}_{l,r})]$  before
computing $Z^{(N)}$. This choice leaves unchanged the weight of loops which do not encircle nor intersect the cut. A loop that surround
{\em both} ends (and thus, lives on a single sheet) gathers ${\rm e}^{\pm i\pi e_0}$ from
the turns, and ${\rm e}^{\pm i\pi (e_l+e_r)}={\rm e}^{\mp 2i\pi e_0}$ from the vertex operators (since the loop increases the height of points $l$ and $r$ by $\pm \pi$).
The two contributions give in the end ${\rm e}^{\mp i\pi e_0}$, summing up to $n$ as required. Finally, for a loop encircling only {\em one} end we get phases
${\rm e}^{\pm ie_{l,r} N\pi } {\rm e}^{\pm iN\pi e_0}= {\rm e}^{\pm i N\pi e}$,
so the correct weight $n$ is obtained setting $e=\tfrac{e_0}{N}$.
%

To evaluate the $Z^{(N)}$ we implement the sewing  conditions on the surface,
$\phi_j(z^+)=\phi_{j+1}(z^-)$ with $j$ mod $N$, by forming combinations of the fields that obey twisted boundary conditions along the cut.
For instance, with $N=2$, we form $\phi_+= (\phi_1+\phi_2)/\sqrt{2}$ and $\phi_-= (\phi_1-\phi_2)/\sqrt{2}$. While $\phi_+$ does not see the cut,
$\phi_-$ is now twisted: $\phi_-(z^+)=-\phi_-(z^-)$. For arbitrary $N$, the field $\phi_{\rm sym} \equiv (\phi_1+\ldots+\phi_N)/\sqrt{N}$ does not see the cut,
while the others are twisted by angles ${\rm e}^{2i\pi k/N}$ with $k=1,\ldots,N-1$. Using that the dimension of the twist fields in a {\sl complex} bosonic theory is \cite{Martinec}
$h_{k/N}=k(N-k)/2N^2$ we find that the twisted contribution to the partition function is 
$Z^{(N)}({\rm twist})\propto L^{-2x_N}$
with $x_N=\sum_{k=1}^{N-1} h_{k/N}= \tfrac{1}{12}\left(N-{1\over N}\right)$.
Meanwhile, the field $\phi_+$, which would not contribute to the EE for a free boson theory (here $e_0=0$), now yields a non-trivial term due
to the vertex operators with $e_{l,r}$:
$Z^{(N)}({\rm charge})\propto L^{-2x'_N}$ with $x'_N=N \tfrac{e^2-e_0^2}{2g}=\tfrac{e_0^2}{2g} \left(\tfrac{1}{N}-N\right)$.
Assembling everything we get $Z^{(N)}\propto L^{-\frac{1}{6}\left(N-\frac{1}{N}\right)(1-6e_0^2/g)}$.
Inserting $e_0={1\over x+1}$ and $g={x\over x+1}$ gives the R\'enyi entropies
\begin{equation}
\tilde{S}_{L}^{(N)}= \tfrac{N+1}{6 N}\left[1-\tfrac{6}{x(x+1)}\right]\ln L
\end{equation}
($\tilde{S}$ is obtained for $N \to 1$), hence  proving our claim.
\begin{figure}
\centering
    \includegraphics[scale=.2]{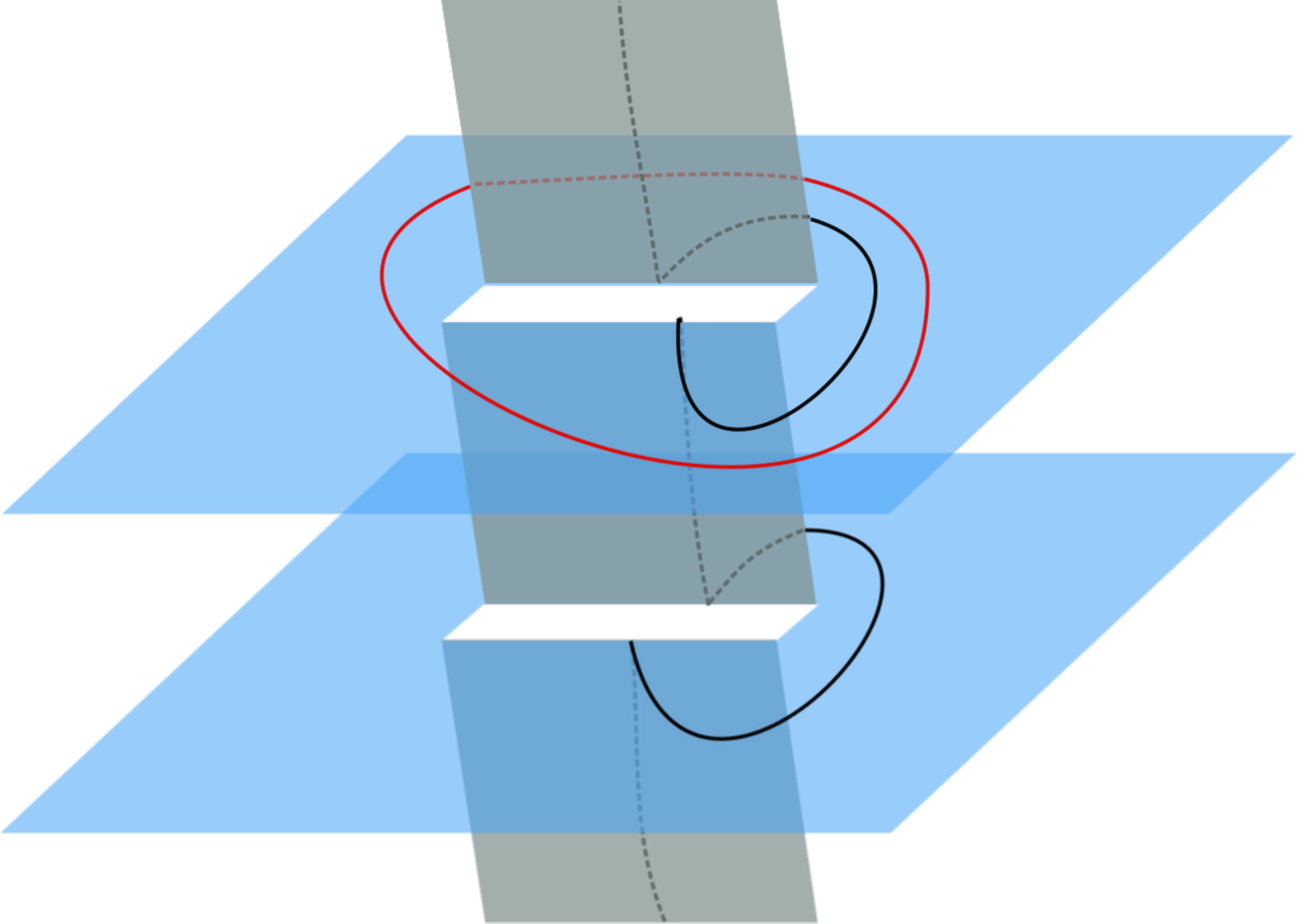}
    \caption{On the Riemann surface used to calculate the Renyi entropy with $N$ replicas (here $N=2$), the black loop must wind $2\pi N$ times before closing onto itself. The red loop surrounds both ends of the cut.}\label{fig1}
\end{figure}

We emphasize that the $U_qsl(2)$ spin chain differs from the usual one simply by the boundary terms $h_i$. These are not expected to affect the ordinary
EE, and  the central charge obtained via the density operator $\rho=|0\rangle\langle 0|$ (with $|0\rangle \propto |0_{\rm R}\rangle$, but normalized as
in our introduction) will be $c_{\rm eff}=1$.

\paragraph{Entanglement in non-unitary minimal models.}
We  now  discuss the restricted solid-on-solid (RSOS) lattice models, which provide the nicest regularization of non-unitary CFTs.
In these models,  the variables are ``heights'' on an $A_m$ Dynkin diagram, with Boltzmann weights that provide yet another
representation of the TL algebra (\ref{TLrels}), with parameter $n=2\cos \tfrac{\pi p}{m+1}$ and $p=1,\ldots,m$. The case $p=1$ is Hermitian,
while $p\neq 1$ leads to negative weights, and hence a non-unitary CFT. One has $c=1-6 {p^2\over (m+1)(m+1-p)}$, and, for $p\neq 1$,
the effective central charge---determined by the state of lowest conformal weight \cite{ISZ} through $c_{\rm eff}=c-24h_{\rm min}$---is
$c_{\rm eff}=1-{6\over (m+1)(m+1-p)}$ . The case $(m,p)=(4,3)$ gives the Yang-Lee singularity universality class discussed in the introduction. 

Defining the EE for RSOS models is not obvious, since their Hilbert space (we use this term even in the non-unitary case) is not a tensor
product like for spin chains. Most recent numerical and analytical work however neglected this fact, and EE was defined using a
straightforward partial trace, summing over all heights in $B$ compatible with those in $A$. In this case, it was argued  and checked numerically
that $S_A= \tfrac{c}{3}\ln L$ in the unitary case, and $S_A = \tfrac{c_{\rm eff}}{3} \ln L$ in the non-unitary case. Note that $c$ matches that of
the loop model based on the same TL algebra, with $x+1\equiv {m+1\over p}$.
For details on the QG EE in the RSOS case, see the SM.

The RSOS partition functions can be expressed in terms of loop model ones, $Z_\ell$.  In the plane, the equivalence \cite{Pasquier} replaces equal-height clusters
by their surrounding loops, which get the usual weight $n$ through an appropriate choice of weights on $A_m$. With periodic boundary conditions,
the correspondence is more intricate due to non-contractible clusters/loops. On the torus \cite{DFSZ1}, $Z_\ell$ is defined by giving each
loop (contractible or not) weight $n$, whereas for the RSOS model contractible loops still have weight $n$, but one sums
over sectors where each non-contractible loop gets the weights $n_k=2\cos \tfrac{\pi k}{m+1}$ for any $k=1,\ldots,m$. The same sum
occurs (see SM for details) when computing $Z^{(N)}$ of the Riemann surface with $N$ replicas: non-contractible loops are here those
winding one end of the cut. Note also that $|0_{\rm L}\rangle=|0_{\rm R}\rangle$
for RSOS models, so the imaginary-time definition of $\rho$ in unambiguous \cite{Doyon1,Doyon2}.

Crucially, the sum over $k$ is dominated (in the scaling limit) by the sector with the largest $n_k$, that is $k=1$ and $n_1=2\cos \tfrac{\pi}{m+1}$.
In the non-unitary case ($p > 1$), $n_1 \neq n$, and the EE is found by extending the above computation. We have still
$e_0= \tfrac{p}{m+1}$, but now $e= \tfrac{1}{N(m+1)} = \tfrac{e_0}{p N}$. To normalize at $N=1$, one must divide by $Z^{(1)}$
to the power $N$, with the same charges:
\begin{equation}
Z^{(N)} / \big( Z^{(1)} \big)^N \propto L^{- \frac{1}{6}\left(N-\frac{1}{N}\right) \big(1-\frac{6e_0^2}{p^2g} \big)} \,, \label{nonunit}
\end{equation}
whence the R\'enyi entropy $S_A^{(N)}=\tfrac{N+1}{6N}c_{\rm eff}\ln L$. 
Hence our construction establishes the claim of \cite{Doyon1,Doyon2}.

\paragraph{EE in the $sl(2|1)$ SUSY chain.}
Percolation and other problems with SUSY (see the introduction) have $Z=1$, hence $c=0$, and the EE scales trivially.
Having a non-trivial quantity that distinguishes the many $c=0$ universality classes would be very
useful. We now show that, by carefully distinguishing left and right eigenstates, and using traces instead of supertraces,
one can modify the definition of EE to build such a quantity.

We illustrate this by the $sl(2|1)$ alternating chain \cite{ReadSaleur01} which describes percolation hulls.
This chain represents the TL algebra (\ref{TLrels}) with $n=1$, and involves the fundamental ($V$) and its conjugate ($\bar{V}$) on alternating sites, with $\dim V = 3$.
The 2-site Hamiltonian, $H=-e_1$, restricted to the subspace $\{ \ket{1\bar{1}},\ket{2\bar{2}},\ket{3\bar{3}} \}$ (where $1,2$ are bosonic and $3$ is fermionic), reads 
\be
 e_1 = |0_{\rm R}\rangle\langle 0_{\rm L}| = \left(\ket{1\bar{1}}+\ket{2\bar{2}}+\ket{3\bar{3}}\right)\left( \bra{1\bar{1}}+\bra{2\bar{2}}+\bra{3\bar{3}}\right) \nonumber
\ee
The eigenvectors are
$|0_{\rm R}\rangle = \ket{1\bar{1}}+\ket{2\bar{2}}+\ket{3\bar{3}}$ and $\langle 0_{\rm L}|= \bra{1\bar{1}}+\bra{2\bar{2}}-\bra{3\bar{3}}$;
note that conjugation is supergroup invariant (i.e., $\langle \bar{3} \ket{\bar{3}}=-1$). Hence, despite the misleading expression, $H$ is not unitary.
The density operator is $\tilde{\rho}=e_1$ and
satisfies $\STr \tilde{\rho} \equiv \Tr (-1)^F\tilde{\rho}= 1$. The reduced density operator 
$\tilde{\rho}_A = \STr_B \tilde{\rho} =  \ket{1}\bra{1}+\ket{2}\bra{2}+\ket{3}\bra{3}$.
If we define the R\'enyi EE also with the supertrace, we get $\STr \tilde{\rho}_A^N=1$ for all $N$.  It is more interesting (and natural) to take instead
the {\sl normal trace} of $\tilde{\rho}$; this requires a renormalization factor to ensure $\Tr\tilde{\rho}_A=1$. We obtain then 
$\tilde{\rho}_A^N=\frac{1}{3^N}\left(\ket{1}\bra{1}+\ket{2}\bra{2}+\ket{3}\bra{3}\right)$ and thus $\tilde{S}_A^{(N)}=\ln 3$.
This equals the QG R\'enyi EE with $n=3$. 

This calculation carries over to arbitrary size. One finds that $\tilde{S}_A = \tilde{S}_{A,\ell}$ with weight $n=1$, {\sl provided} non-contractible
loops winding around one cut end in the replica calculation get the modified weight $\tilde{n}=3$ instead of $n$. We can then use the CG framework
developed in the context of the non-unitary minimal models to calculate the scaling behavior. We use  
(\ref{nonunit}), with $g=\tfrac{2}{3}$ for percolation ($n=1$), and $\tilde{n}=2\cos\pi e_0$. It follows that $e_0$ is purely imaginary,
and that  $\tilde{S}^{(N)}\thicksim \tfrac{N+1}{6 N}c_{\rm eff}\log L$
with $ c_{\rm eff} = 1+\frac{9}{\pi^2}\big(\log\frac{3+\sqrt{5}}{2}\big)^2\thicksim1.84464\ldots$. 

\paragraph{Numerical checks.} All these results were checked numerically. As an illustration, we discuss only the case 
$q={\rm e}^{2i\pi/5}$, for which the RSOS and loop models have $c=-3/5$, while $c_{\rm eff}=3/5$ for the RSOS model.
In the corresponding $U_qsl(2)$ chain, we measured the (ordinary) EE as in (\ref{Svn}), the QG R\'enyi EE $\tilde{S}^{(2)}$
as in (\ref{Svnmod}), and the QG R\'enyi EE for the modified loop model where non-contractible loops have
fugacity $n_1=2\cos{\pi\over 5}$ (instead of $n=2\cos \frac{2 \pi}{5}$). This, recall, should coincide asymptotically with
the R\'enyi EE for the RSOS model. Results (see figure~\ref{fig1}) fully agree with our predictions. 

\begin{figure}
\centering
    \includegraphics[scale=.7]{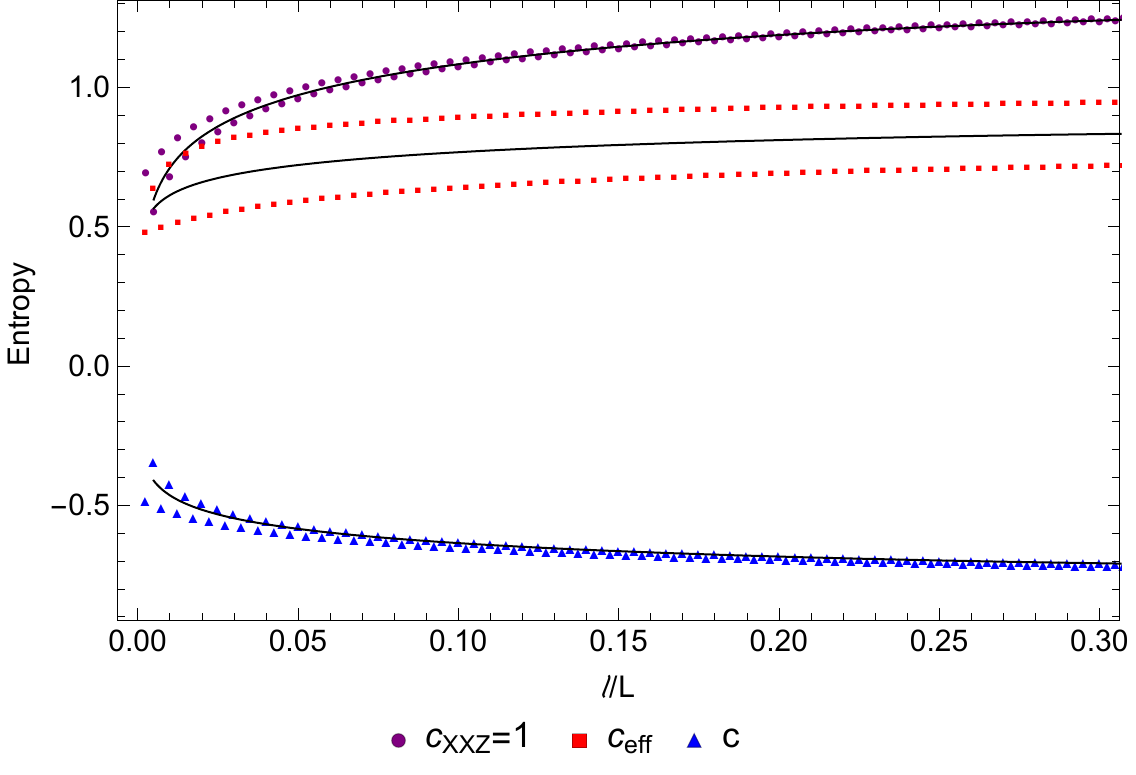}
    \caption{Numerical EE for the non-unitary case $(p,m)=(2,4)$ ($n=2\cos \frac{2\pi}{5}$), versus the length of the cut $L$, for a chain with $M=400$ sites and open boundary conditions. Purple dots show the usual EE with the unmodified trace. Averaging over the parity oscillations (solid curve) reveals the scaling with $c_{\rm XXZ}=1$. Red squares show the $N=2$ R\'enyi entropy, with the modified trace giving weight $n_1=2\cos\pi/5$ to non-contractible loops; this scales with $c_{\rm eff}=3/5$. Blue triangles again show $\tilde{S}^{(2)}$, but with $n_1=n$; the scaling then involves the true central charge $c=-3/5$.}\label{fig1}
\end{figure}

\paragraph{\sl Conclusion.}  While we have mostly discussed the critical case, we stress that the QG EE
can be defined also away from criticality. An interesting example is the $sl(2|1)$ alternating chain,
for which staggering makes the theory massive (this corresponds to shifting the topological angle away from $\Theta=\pi$
in the sigma-model representation). Properties of the QG R\'enyi EE along this (and other) RG flows will be reported elsewhere. 

To summarize, we believe that our analysis completes our understanding of EE in 1D by providing a natural extension to
non-unitary models in their critical or near-critical regimes. There are clearly many situations (such as phenomenological
``Hamiltonians'' for open systems) where things will be very different, but we hope our work will provide the first step in the
right direction. Our approach also provides a long awaited ``Coulomb gas'' handle on the correspondence between lattice
models and quantum information quantities. In the SM we apply this to show that, in the case of non-compact theories,
the well-known ${c\over 3}\ln L$ term will be corrected by $\ln\ln L$ terms (with, most likely, a non-universal amplitude),
in agreement with recent independent work \cite{BenjaminOlalla}.

\noindent{\sl Acknowledgments}: The work of HS and JLJ was supported by the ERC Advanced Grant NuQFT. The work of HS was also supported by the US Department of Energy (grant number DE-FG03-01ER45908). We thank B. Doyon and O. Castro-Alvaredo for inspiring discussions and comments.

\vfill
\eject

\clearpage
\newpage

\section{Supplementary Material}

In these notes we provide additional details for some results of the main text. We first provide additional motivation for our definition of the entanglement entropy (EE) from
the perspective of the $U_qsl(2)$ quantum group (QG) symmetry, and we prove that $\tilde{S}_A = \tilde{S}_B$. Next, we give more examples of the computation of the QG
EE for larger larger systems in various representations. We elaborate on the construction in the RSOS case, detailing in particular the mapping between the RSOS and
loop model representations. Finally we discuss the emergence of a $\ln(\ln L)$ term in the non-compact case.

\subsection{$U_qsl(2)$ symmetry for the reduced density operator}

Our definition of the EE relies on using a modified trace, known as a Jones trace, in which a factor of the type $q^{-2S \sigma^z_B}$ is inserted under the usual trace symbol.
To ensure that the resulting reduced density operator $\rho_A$ makes sense in the QG formalism, we must ensure that it commutes with the generators of $U_qsl(2)$.

We therefore consider the XXZ spin-$1/2$ chain, with boundary terms as described in the main text. The Hamiltonian commutes with the following generators:
\small
\bea  S^z&=&\sum_i \sigma_i^z\\
S^+&=&\frac{1}{2}\sum_i q^{\sigma_1^z}\! \otimes\!...\!\otimes\! q^{\sigma_{i-1}^z}\!\otimes\!\sigma_i^+\!\otimes\! q^{-\sigma_{i+1}^z}\!\otimes\!...\!\otimes\! q^{-\sigma_{M}^z}\\
S^-&=&\frac{1}{2}\sum_i q^{\sigma_1^z} \!\otimes\!...\!\otimes\! q^{\sigma_{i-1}^z}\!\otimes\!\sigma_i^-\!\otimes\! q^{-\sigma_{i+1}^z}\!\otimes\!...\!\otimes\! q^{-\sigma_{M}^z}\eea
\normalsize
Since the generators commute with the Hamiltonian, they share the same right and left eigenvectors. As a consequence they commute with the density operator $\rho$
\be [S^\alpha,\rho]=0,\qquad\rho = \ket{0_{\rm R}}\bra{0_{\rm L}}\,. \ee
We split the spin chain in two parts $A$, $B$ and define the reduced density operator $\rho_A$ using a Jones trace over the part $B$. We consider the case where $A$ is in the middle of the chain between $B_{\rm L}$ and $B_{\rm R}$, so that $B = B_{\rm L} \cup B_{\rm R}$ and $\mathcal{H}=\mathcal{H}_{B_{\rm L}}\otimes\mathcal{H}_{A}\otimes\mathcal{H}_{B_{\rm R}}$. Thus
\small
\be \rho_A=\Tr_B\ q^{2S^z_{B_{\rm L}}-2S^z_{B_{\rm R}}}\rho \,, \quad \mbox{with } S^z_B=\sum_{i\in B}\sigma_i^z \,. \ee
\normalsize
Let us check that the generators of $U_q(sl_2)$ on the subsystem $A$ commute with the reduced density operator $\rho_A$. We have the following relations:
\bea S^z\!&=&\!S^z_{B_{\rm L}}\otimes\mathds{1}+\mathds{1}\otimes S^z_A\otimes\mathds{1} + \mathds{1}\otimes S^z_{B_{\rm R}} \,, \\
S^\pm\!&=&\!S^\pm_{B_{\rm L}} q^{-S^z_A-S^z_{B_{\rm R}}}\!+\!q^{S^z_{B_{\rm L}}} S^\pm_A q^{-S^z_{B_{\rm R}}} \!+\! q^{S^z_{B_{\rm L}}+S^z_A}S^\pm_{B_{\rm R}} \,. \nonumber
\eea
\normalsize
Consider first $S^z_A$:
\small
\bea S^z_A\ \rho_A &=& \Tr_B\left(S^z_Aq^{2S^z_{B_{\rm L}}-2S^z_{B_{\rm R}}}\rho\right) \\
&=& \Tr_B\left((S^z-S^z_{B_{\rm L}}-S^z_{B_{\rm R}})q^{2S^z_{B_{\rm L}}-2S^z_{B_{\rm R}}}\rho\right) \,. \nonumber \eea
\normalsize
Obviously $S^z_{B_{\rm L}}$, $S^z_{B_{\rm R}}$, $S^z_A$ and $S^z$ commute. Since $S^z$ also commutes with $\rho$:
\small
\bea S^z_A\ \rho_A &=& \Tr_B\left(q^{2S^z_{B_{\rm L}}-2S^z_{B_{\rm R}}}\rho S^z\right)\nonumber- \Tr_B\left(q^{2S^z_{B_{\rm L}}-2S^z_{B_{\rm R}}}\rho S^z_{B_{\rm L}}\right)\\&-&\Tr_B\left(q^{2S^z_{B_{\rm L}}-2S^z_{B_{\rm L}}}\rho S^z_{B_{\rm L}}\right) \,. \nonumber\eea
\normalsize
For the two last terms we performed a cyclic permutation under the trace. We can now sum all terms and this proves $\left[S^z_A,\rho_A\right]=0$.
Next we do the same for $S^+_A$:
\small
\bea S^+_A\ \rho_A &=& \Tr_B\left(S^+_A\rho\ q^{2S^z_{B_{\rm L}}-2S^z_{B_{\rm R}}}\right)\nonumber\\
&=&\Tr_B\Big((q^{-S^z_{B_{\rm L}}+S^z_{B_{\rm R}}}S^+-q^{-S^z_A-S^z_{B_{\rm L}}}S^+_{B_{\rm L}}\nonumber\\
&&-q^{S^z_A+S^z_{B_{\rm R}}}S^+_{B_{\rm R}})\rho q^{2S^z_{B_{\rm L}}-2S^z_{B_{\rm R}}}\Big)\nonumber\\&\equiv&(1)-(2)-(3) \,. \eea
\normalsize
The first term $(1)$ of the right-hand side reads
\small
\bea
(1)&=&\Tr_B\Big(q^{-S^z_{B_{\rm L}}+S^z_{B_{\rm R}}}S^+\rho q^{2S^z_{B_{\rm L}}-2S^z_{B_{\rm R}}}\Big)\nonumber\\&=&\Tr_B\Big(q^{2S^z_{B_{\rm L}}-2S^z_{B_{\rm R}}}\rho S^+q^{-S^z_{B_{\rm L}}+S^z_{B_{\rm R}}}\Big)\nonumber
\eea
\normalsize
thanks to the cyclic permutation under the trace and the commutation of $\rho$ and $S^+$. We then deal with the second term $(2)$ involving $S^+_{B_{\rm L}}$:
\small
\bea
(2)&=&\Tr_B\Big(q^{-S^z_A-S^z_{B_{\rm L}}}S^+_{B_{\rm L}}\rho q^{2S^z_{B_{\rm L}}-2S^z_{B_{\rm R}}}\Big)\nonumber\\
&=&\Tr_B\Big(S^+_{B_{\rm L}} q^{-S^z_A} \rho q^{S^z_{B_{\rm L}}-2S^z_{B_{\rm R}}} \Big)\nonumber\\
&=&\Tr_B\Big(q^{-S^z+S^z_{B_{\rm R}}+S^z_{B_{\rm L}}} \rho q^{S^z_{B_{\rm L}}-2S^z_{B_{\rm R}}} S^+_{B_{\rm L}}\Big)\nonumber\\
&=&\Tr_B\Big(q^{-S^z_{B_{\rm R}}+S^z_{B_{\rm L}}} \rho q^{-S^z+S^z_{B_{\rm L}}} S^+_{B_{\rm L}}\Big)\nonumber\\
&=&\Tr_B\Big(q^{-2S^z_{B_{\rm R}}+2S^z_{B_{\rm L}}} \rho S^+_{B_{\rm L}}q^{-S^z_A-S^z_{B_{\rm L}}} \Big) \,, \nonumber
\eea
\normalsize
thanks to cyclic permutations of the operators over the subsystem $B$, the commutation of $S^z$ with $\rho$ and the commutation of $S^z_{B_{\rm L}}$ with $S^z_A$ and $S^z_{B_{\rm R}}$. Similarly for the term $(3)$ involving $S^+_{B_{\rm R}}$:
\small
\bea
(3)&=&\text{Tr}_B\Big(q^{S^z_A+S^z_{B_{\rm R}}}S^+_{B_{\rm R}}\rho q^{2S^z_{B_{\rm L}}-2S^z_{B_{\rm R}}}\Big)\nonumber\\
&=&\text{Tr}_B\Big(q^{-2S^z_{B_{\rm R}}+2S^z_{B_{\rm L}}} \rho S^+_{B_{\rm R}}q^{S^z_A+S^z_{B_{\rm R}}} \Big) \,. \nonumber
\eea
\normalsize
By regrouping the terms we find the desired property $S^+_A\rho_A=\rho_AS^+_A$. A very similar computation can be done for $S^-_A$.

\subsection{Proof of $\tilde{S}_A=\tilde{S}_B$}

A meaningful EE must satisfy, at the very least, the symmetry property $S_A = S_B$, meaning that subsystem $A$ is as entangled with $B$, as $B$ with $A$.
We now show that this is the case for our QG EE.

Let us consider the case $q \in \mathbb{R}$. The proof is then simple and can be extended by analytic continuation to complex $q$. In this case the Hamiltonian is symmetric, and $\ket{0}\equiv\ket{0_{\rm R}}=\ket{0_{\rm L}}$. We again divide our system in two pieces $A$ and $B$ with a cut in the middle (for more complicated cuts the argument is similar) and write the state in the following way:
\bea \ket{0}=\sum_{i,j}\psi_{i,j}\ket{i}_A\ket{j}_B \,. \eea
The bases $\ket{i}_A$ and $\ket{j}_B$ can be chosen such that they have a well-defined magnetization. As a consequence, since the groundstate $|0\rangle$ is in the zero-magnetization sector, we can define those bases such that the matrix $\psi_{i,j}$ is block-diagonal and where each block corresponds to a sector of $A$ and $B$ with a well-defined magnetization. When we perform a singular value decomposition (SVD) we end up with
\bea \ket{0}= \sum_{\alpha}s_\alpha\ket{\alpha}_A\ket{\alpha}_B \,, \eea
where $\ket{\alpha}_A$ and $\ket{\alpha}_B$ are eigenvectors of $S_A$ and $S_B$; they form orthonormal bases of $A$ and $B$. The density matrix $\rho$ is 
\bea \rho = \sum_{\alpha,\alpha'} s_\alpha s_{\alpha'} \ket{\alpha}_A\ket{\alpha}_B\bra{\alpha'}_A\bra{\alpha'}_B \,.
\eea
The reduced density matrices $\rho_A$ and $\rho_B$ read
\bea \rho_A &=& \Tr_Bq^{-2S_B}\rho=\sum_{\alpha} s_\alpha^2 q^{-2S^\alpha_B} \ket{\alpha}_A\bra{\alpha}_A \,, \nonumber\\
\rho_B &=& \Tr_Aq^{-2S_A}\rho=\sum_{\alpha} s_\alpha^2 q^{2S^\alpha_A} \ket{\alpha}_B\bra{\alpha}_B \,. \nonumber
\eea
Since the ground state is in the $S=0$ sector $q^{2S^\alpha_A}=q^{-2S^\alpha_B}$ and thus the two reduced density operators have the same spectra and define the same entropy. This proves the statement $\tilde{S}_A=\tilde{S}_B$ in the case of a cut in the middle of the system.

\subsection{More examples}

To keep the discussion in the main text as simple as possible, we have presented all explicit computations for a chain with just $M=2$ sites.
This is of course no limitation to applying our general definitions, and accordingly we give here a few examples for higher values of $M$.

\subsubsection{Loop representation}

We consider the case of $M=4$ sites. The basis of link states is : 
\be \ket{1} = \linkStateA,\qquad\ket{2} = \linkStateB \ee
\\ 
The hamiltonian $H=-e_1-e_2-e_3$ has the following ground state $\ket{0}=\frac{1}{{\cal N}}(\alpha\ket{1}+\ket{2})$,
where ${\cal N}^2=n^2\alpha^2+2n\alpha+n^2$ and $\alpha = (n+\sqrt{n^2+8})/2$.
The density matrix $\rho$ is
\be
\rho = \frac{1}{{\cal N}^2}\Big(\alpha^2 \!\!\!\!
\raisebox{0.6em}{\wordA}\!\!\!\!
+\alpha\!\!\!\!\raisebox{0.6em}{\wordB}\!\!\!\!
+\alpha\!\!\!\!\raisebox{0.6em}{\wordC}\!\!\!\!
+\!\!\!\!\raisebox{0.6em}{\wordD}\!\!\!\!
\Big) \,. \nonumber
\ee

Consider first a bipartition in which $A$ is the first site, and $B$ the remainder.
Take the partial Markov trace over the three last sites, we find the reduced density operator
\be
\rho_A=\frac{1}{{\cal N}^2}(\alpha^2n+2\alpha+n)\!\!\!\!\ReducedDensityMatOneSite\!\!\!\!=\frac{1}{n}\!\!\!\!\ReducedDensityMatOneSite \,. \ee
This leads to $\tilde{S}_A=\log n$, the same result as found in the main text for the EE of the first spin with $M=2$.

Next we take $A$ as the first two sites, to compute the entanglement at the middle of the system. 
We trace the density operator over the two last sites:
\bea
\rho_A &=& \frac{1}{{\cal N}^2}\Big( \ \IdentityTwoSites
+ (\alpha^2n+2\alpha)\DensityMatTwoSites \Big) \nonumber \\ 
&=& \frac{1}{{\cal N}^2} \big( \mathbb{I} + (\alpha^2 n + 2 \alpha) e_1 \big) \,. \eea
We now need to take the logarithm of $\rho_A$.
We notice the identity $\exp(a\,e_1)=1+\frac{1}{n}(\exp(an)-1)e_1$, where $e_1=\TLGenInline$.
It is then easy to find that \be\log\rho_A = -\log {\cal N}^2 \, \mathbb{I} +\frac{2}{n}\log(1+\alpha n) e_1 \,. \label{logrhoAlemma} \ee
We can now compute $\rho_A\log\rho_A$ as
\small
\bea &-&\frac{\log {\cal N}^2}{{\cal N}^2} \ \IdentityTwoSites +
\Big(\frac{-\log{{\cal N}^2}}{{\cal N}^2}(\alpha^2n + 2\alpha)\nonumber + \frac{2}{n{\cal N}^2}\log(1 + \alpha n) \\
&+&\frac{2}{{\cal N}^2}(\alpha^2n+2\alpha)\log(1+\alpha n)\Big) \DensityMatTwoSites \,.
\eea
\normalsize
Tracing over $A$ we finally obtain
\small
\bea\nonumber \tilde{S}_A &=&-\Tr \rho_A\log\rho_A = \frac{n^2\log {\cal N}^2}{{\cal N}^2} - n\Big(\frac{-\log{{\cal N}^2}}{{\cal N}^2}(\alpha^2n+2\alpha)\\\nonumber &+&\frac{2}{n{\cal N}^2}\log(1+\alpha n)+\frac{2}{{\cal N}^2}(\alpha^2n+2\alpha)\log(1+\alpha n)\Big) \\
&=& - \frac{(1 + \alpha n)^2}{{\cal N}^2} \log (1 + \alpha n)^2 + \log {\cal N}^2 \,. \label{S4sites}
\eea
\normalsize
We have verified that this expression coincides with the result obtained by using the modified trace in the vertex model.
It also agrees with computations in the Potts spin representation for $Q=n^2$ integer (see below).

For larger $M$ it is hard to compute this final partial trace directly, since the form of $\log \rho_A$ will be substantially more complicated than (\ref{logrhoAlemma}).
A much more convenient option is to recall that gluing corresponding sites on top and bottom of any word in the TL algebra means technically to take the so-called
Markov trace $\MTr$. This in turn can be resolved as follows
\be \MTr \  = \sum_j [2j+1]_q \Tr_{{\cal V}_j} \,, \label{Markovtracedecomp} \ee
where $\Tr_{{\cal V}_j}$ is the usual matrix trace over the (standard) module ${\cal V}_j$ with $2j$ defect lines, and $[k]_q = \tfrac{q^k - q^{-k}}{q-q^{-1}}$ are $q$-deformed numbers
such that the loop weight $n = [2]_q = q+q^{-1}$.

In the simple $M=4$ case considered above, $A$ has just two sites so that
${\cal V}_0$ and ${\cal V}_1$ are both one-dimensional with bases $\{ \gsNtwoInline \}$ and $\{ \IdentityTwoSitesInline \}$ respectively.
Thus we have the matrices
\be \left. \rho_A \right|_{{\cal V}_0}= \left[ \frac{1}{{\cal N}^2}(1+ n\ (\alpha^2 n+2\alpha)) \right] \,,
\quad \left. \rho_A \right|_{{\cal V}_2}= \left[ \frac{1}{{\cal N}^2} \right]  \nonumber \ee
and
\bea \MTr \rho_A\log\rho_A &=& \Tr_{{\cal V}_0}\rho_A\log\rho_A \nonumber \\
 &+& (n^2-1) \Tr_{{\cal V}_2}\rho_A\log\rho_A \,. \label{Markovtrace1} \eea
We find in the end
\bea
\MTr\rho_A\log\rho_A &=& \frac{(1+ \alpha n)^2}{N^2} \log\frac{(1+\alpha n)^2}{{\cal N}^2}\nonumber\\&+&(n^2-1)\frac{1}{{\cal N}^2}\log{\frac{1}{{\cal N}^2}} \,, \eea
which is the same as (\ref{S4sites}) after simplification.

We have made similar computations for $M=6$ sites, for all choises of the bipartition $A \cup B$, finding again perfect agreement
between the results from the loop model (with the Markov trace) and the vertex model (with the modified trace). 

\subsubsection{Spin representation}

The same computation can be conducted in the $Q$-state Potts spin representation for $Q$ integer. There are ${\cal L}=M/2$ spins labelled $S_j$ (with $j=1,2,\ldots,{\cal L}$)
enjoying free boundary  conditions. The interactions $e_i$ take different expressions depending
on the parity of $i$. We have $e_{2j-1} = Q^{-1/2} D_j$, where $D_j$ detaches the $j$'th spin from the rest (the new spin freely takes any of the $Q$ values);
while $e_{2j} = Q^{1/2} J_j$, where $J_j = \delta(S_j,S_{j+1})$ joins two neighbouring spins (forcing them to take the same value).

In the above $M=4$ example, the Hamiltonian is
\be
 H = - Q^{-1/2}(D_1+D_2) - Q^{1/2} J_1 \,,
 \label{HIsing}
\ee
and for $Q=2$ the interactions read explicitly, in the basis $\left\lbrace |++\rangle, |+-\rangle, |-+\rangle, |--\rangle \right\rbrace$,
\be
 D_1 = \left( \begin{smallmatrix}
 1 & 0 & 1 & 0 \\
 0 & 1 & 0 & 1 \\
 1 & 0 & 1 & 0 \\
 0 & 1 & 0 & 1 \\ \end{smallmatrix} \right) \,, \quad
 D_2 = \left( \begin{smallmatrix}
 1 & 1 & 0 & 0 \\
 1 & 1 & 0 & 0 \\
 0 & 0 & 1 & 1 \\
 0 & 0 & 1 & 1 \\ \end{smallmatrix} \right) \,, \quad
 J_1 = \left( \begin{smallmatrix}
 1 & 0 & 0 & 0 \\
 0 & 0 & 0 & 0 \\
 0 & 0 & 0 & 0 \\
 0 & 0 & 0 & 1 \\ \end{smallmatrix} \right) \,. \nonumber
\ee
The normalised ground state is
\be
 |0\rangle = k^- \left( |++\rangle + |--\rangle \right) + k^+ \left( |+-\rangle + |-+\rangle \right) \,,
\ee
with $k^\pm = (5 \pm \sqrt{5})^{-1/2}$. Tracing over the subsystem $B$ (the right spin $S_2$) we find the reduced density matrix
\be
 \rho_A = \left( \begin{array}{cc} (k^+)^2 + (k^-)^2 & 2 k^+ k^- \\ 2 k^+ k^- & (k^+)^2 + (k^-)^2 \end{array} \right)
 = \left( \begin{array}{cc} \frac{1}{2} & \frac{1}{\sqrt{5}} \\ \frac{1}{\sqrt{5}} & \frac12 \end{array} \right) \,, \nonumber
\ee
where of course $\Tr \rho_A = 1$. The eigenvalues are $\lambda^\pm = \frac12 \pm \frac{1}{\sqrt5}$. Thus
\bea
 \tilde{S}_A &=& -\left( \lambda^+ \log \lambda^+ + \lambda^- \log \lambda^- \right) \nonumber \\
 &=& \log \left( 2 \sqrt{5} \right) + \frac{1}{\sqrt{5}} \log \left( 9 - 4 \sqrt{5} \right) \,,
\eea
which is easily seen to agree with (\ref{S4sites}) for $Q=2$.

Let us note that the eigenenergies of (\ref{HIsing}) are $-(3 \pm \sqrt{5})/\sqrt{2}, -2 \sqrt{2}, -\sqrt{2}$. The first two (and in particular the ground state energy)
are also found in the loop model, but the latter two are not. As we have seen, this does not prevent us from finding the same $S$, which is a property of $|0\rangle$.
On the other hand, one can check that $\rho_A$ has the same spectrum in the two representations. These conclusions extend to $Q=3$: we find the same $\tilde{S}_A$,
and the eigenvalues of $\rho_A$ are the same (up to multiplicities, and after the elimination of non-relevant zero eigenvalues). 

\subsubsection{RSOS representation}

In the RSOS construction a height $h_i = 1,2,\ldots,m$ is defined at each site $i=0,1,\ldots,M$, subject to the
constraint $|h_i-h_{i-1}| = 1$ for each $1 < i \le M$.
We note that while the loop model is defined on $M$ strands, there are now $M+1$ RSOS heights.

Free boundary conditions for the first and last spins in the equivalent
Potts model (i.e., no defect lines in the loop model) correspond \cite{BauerSaleur89} to fixing $h_0=h_M=1$. More generally,
having $2j$ defect lines in the loop model would correspond to $h_0=1$ and $h_M=1+2j$.

To explain the details we move to a slightly larger example, namely $M=6$  and $m=5$ (i.e., $Q=3$), in order to see all non-trivial features of
the computation at work. Consider the following labelling of the RSOS basis states:
\begin{eqnarray}
 |1\rangle &=& \{1,2,3,2,3,2,1\} \,, \quad
 |2\rangle = \{1,2,3,2,1,2,1\} \,, \nonumber \\
 |3\rangle &=& \{1,2,3,4,3,2,1\} \,, \quad
 |4\rangle = \{1,2,1,2,3,2,1\} \,, \nonumber \\
 |5\rangle &=& \{1,2,1,2,1,2,1\} \,.
\end{eqnarray}
It is straightforward to find the normalised ground state $|0\rangle$ in this basis and check that its eigenenergy coincides with that of the other representations.
We denote $\rho = |0\rangle \langle 0|$ as usual.

Consider first the bipartion ``4+2'', where $A$ contains the first four sites and $B$ the last two. The junction of the two intervals is at $i^* = 4$, and we
write $h_{i^*} = 1+2j$ for the corresponding intermediate height which belongs to both $A$ and $B$.
The sector $j=0$ has the basis $\{ |2\rangle, |5\rangle \}$, while $j=1$ has the basis $\{ |1\rangle, |3\rangle, |4\rangle \}$.
In each sector, the reduced density matrix $\left. \rho_A \right|_{{\cal V}_j}$ is formed by tracing over the heights belonging to $B$. However,
in both cases the choice of boundary conditions ($h_M=1$), the sector label $j$ and the RSOS constraint fully fix the $B$-heights,
so the trace is trivial:
\bea
 \left. \rho_A \right|_{{\cal V}_0} &=& \left( \begin{array}{cc} \langle 2 | \rho | 2 \rangle & \langle 2 | \rho | 5 \rangle \\
 \langle 5 | \rho | 2 \rangle & \langle 5 | \rho | 5 \rangle \end{array} \right) \,, \nonumber \\
 \left. \rho_A \right|_{{\cal V}_1} &=& \left( \begin{array}{ccc} \langle 1 | \rho | 1 \rangle & \langle 1 | \rho | 3 \rangle & \langle 1 | \rho | 4 \rangle \\
 \langle 3 | \rho | 1 \rangle & \langle 3 | \rho | 3 \rangle & \langle 3 | \rho | 4 \rangle \\
 \langle 4 | \rho | 1 \rangle & \langle 4 | \rho | 3 \rangle & \langle 4 | \rho | 4 \rangle \\ \end{array} \right) \,.
\eea
Each $\left. \rho_A \right|_{{\cal V}_j}$ has precisely one non-zero eigenvalue, $\Lambda_j$. Applying the normalisation
\be
 \lambda_j = \Lambda_j / [1+2j]_q
\ee
we find that they agree with the eigenvalues of $\rho_A$ restrained to the standard module ${\cal V}_j$ with $2j$ defect lines
in the corresponding loop model computation. Therefore $S$ can be computed by the decomposition (\ref{Markovtracedecomp}) of the Markov trace,
which writes explicitly as (\ref{Markovtrace1}) in this simple case.

To see a non-trivial trace over the $B$-system, we consider the same example but with the ``3+3'' bipartition (i.e., $i^* = 3$).
The sector $j=\frac12$ has the basis (for the $A$-heights) $\{ |1,2,1,2\rangle , |1,2,3,2 \rangle \}$. The first basis element corresponds
to states $|5\rangle$ and $|4\rangle$ for the full system, while the second basis elements corresponds to $|1\rangle$ and $|2\rangle$.
Therefore, to form $\rho_A \left(\frac12\right)$ we must sum over those possibilities (which corresponds to tracing over the free height $h_5 = 1, 3$):
\be
 \rho_A\left(\frac12\right) = \left( \begin{array}{cc}
 \langle 5 | \rho | 5 \rangle + \langle 4 | \rho | 4 \rangle & \langle 5 | \rho | 2 \rangle + \langle 4 | \rho | 1 \rangle \\
 \langle 2 | \rho | 5 \rangle + \langle 1 | \rho | 4 \rangle & \langle 2 | \rho | 2 \rangle + \langle 1 | \rho | 1 \rangle \\ \end{array} \right) \,.
\ee
The remainder of the computation proceeds as outlined above, and the end result again agrees with that of the loop model.

\subsection{Extracting the real central charge in the non-unitary case}

We  describe now the general construction of a modified trace in the RSOS models that will enable us
to extract the true central charge from the entanglement, even in the non-unitary case. 
We thus return to the  $A_m$ RSOS model with $n = 2 \cos \frac{\pi p}{m+1}$ and ${\rm gcd}(p,m+1) = 1$.

Set $S_h = \sin \frac{\pi p h}{m+1}$. The interactions $e_i$ satisfying (\ref{TLrels}) propagate $h_i$ into $h'_i$ and read
$e_i = \delta_{h_{i-1},h_{i+1}} \sqrt{S_{h_i} S_{h'_i}} / S_{h_{i-1}}$.
With the boundary conditions $h_0 = h_M = 1$, the ground state $|0\rangle$ then has the same
energy as in the other representations---this is also true in the non-unitary cases $p>1$, provided we resolve the square root as
$\sqrt{S_{h_i} S_{h'_i}} = S_{h_i}$ when $S_{h_i} = S_{h'_i} < 0$.

Obtaining the reduced density matrix $\rho_A$ for a bipartition $A \cup B$ involves a subtle manipulation of the height $h_{i^*}$ situated at the
junction between $A$ and $B$. For each fixed $h_{i^*} \equiv 1+2j$, define $\rho_A(j)$ as the usual trace of $\rho = |0 \rangle \langle 0|$ over the $B$-heights
($h_i$ with $i > i^*$). Thus $\rho_A(j)$ is a matrix indexed by the $A$-heights ($h_i$ with $i < i^*$). The label $j$ is the quantum group spin
of the sector $\rho_A(h_j)$ of the reduced density matrix, and corresponds to having $2j$ defect lines in the loop model
computation. Now let $\{ \Lambda_k(j) \}$ denote the set eigenvalues of $\rho_A(j)$. We claim that $\lambda_k(j) = \Lambda_k(j) / [1+2j]_q$
yield precisely the corresponding loop model eigenvalues (disregarding any zero eigenvalues), and that the
QG entropy $\tilde{S} = - \MTr \rho_A \log \rho_A$ can be constructed
therefrom by computing the Markov trace (\ref{Markovtracedecomp}) over $A$ in the same way as for the loop model.

Note that this implies the following relation
\begin{eqnarray}
&-&\sum_{j,k} [2j+1]_q \lambda_k(j)\ln\lambda_k(j)=  \\
&-&\sum_{j,k}\Lambda_k(j)\ln\Lambda_k(j) + \sum_{j,k}\Lambda_k(j)\ln[2j+1]_q \,, \nonumber
\end{eqnarray}
where on the left we have the QG EE, and the first term on the right is the `ordinary' EE for the (non-unitary) RSOS model.
The term on the left scales like $\frac{c}{6} \ln L$,
and the first term on the right like $\frac{c_{\rm eff}}{6} \ln L$. This implies that the second term on the right must also be proportional to $\ln L$ in the non-unitary case.
While this is not impossible in view of our knowledge of entanglement spectra \cite{LefCal}, the result clearly deserves a more thorough study. 

\subsection{The detailed calculation in the RSOS case}

\begin{figure}
\centering
\includegraphics[scale=.25]{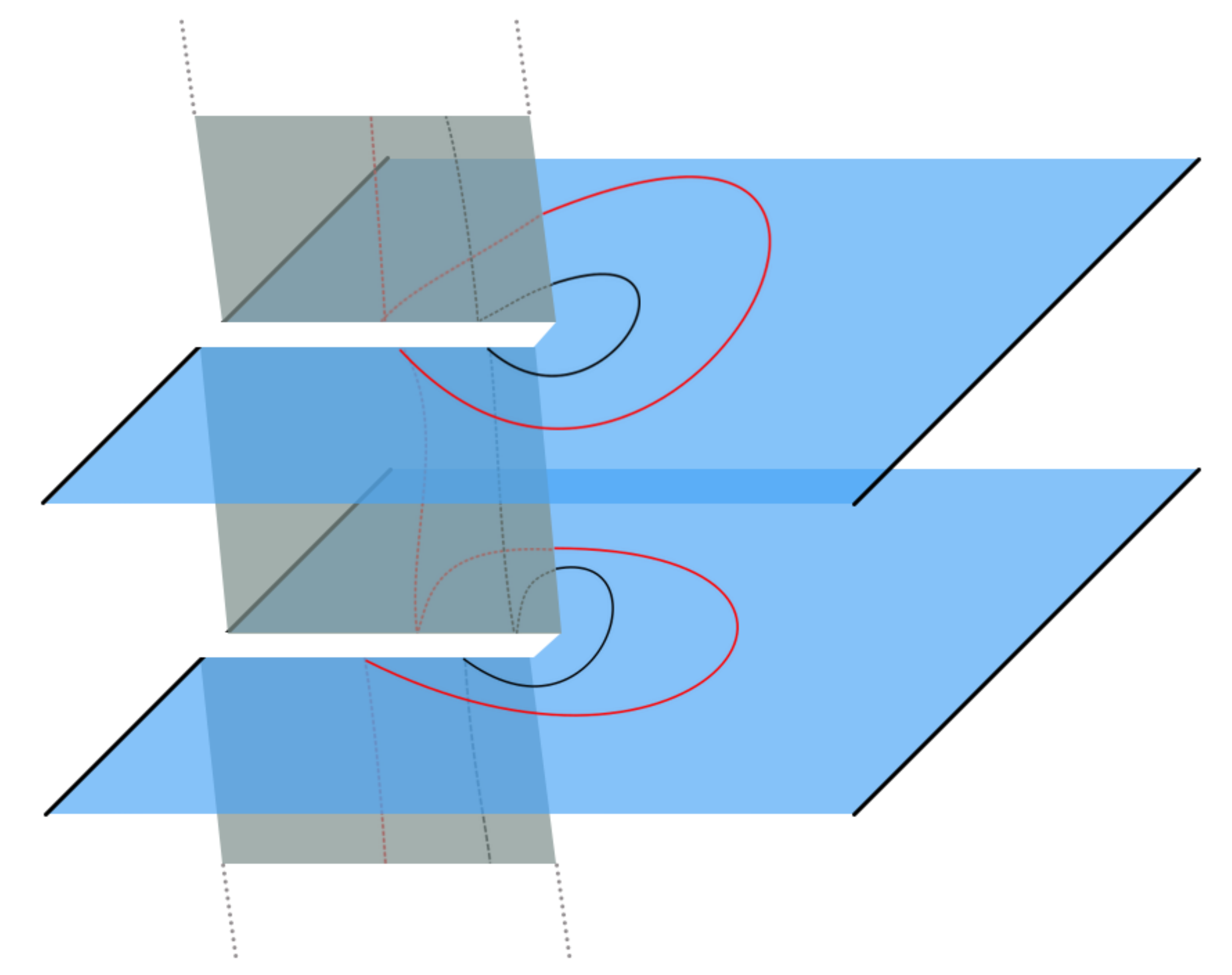}
\caption{Surface with $N=2$ replicas. Black edges along the planes represent boundary conditions (height fixed to $\hbdy$).
The top and bottom are identified. Non-contractible loops always wind $N$ times. Moreover, only the outermost loop (red here)
sees the boundary directly.}\label{loopR}
\end{figure}

We discuss here in more detail the correspondence between the RSOS and loop models for the calculation of the R\'enyi entropies. 
For simplicity, we only consider open boundary conditions with the boundary heights $\hbdy$ fixed to $1$ and a cut on the edge of the system (Figure \ref{loopR}). Loops surround clusters of constant height. When a loop makes a right (resp.\ left) turn by bouncing off a piece of a cluster, it gets a weight $\sqrt{S_a/S_b}$ (resp.\ $\sqrt{S_b/S_a}$) where $a$ and $b$ are the heights of the adjacent clusters (cluster of height $b$ on the left, and $a$ on the right). The amplitude $S_h$ is defined by $S_h=\frac{\sin h\lambda}{\sin \lambda}$, with $\lambda = \frac{p\pi}{m+1}$. After summing over all possible heights, loops pick a weight $n=2\cos\frac{p\pi}{m+1}=\frac{S_{h-1}+S_{h+1}}{S_h}$ if they are homotopic to a point.

Let us consider the R\'enyi entropies for $N>1$. Using the replica picture, we must compute the weigh of loops on the $N$-sheeted surface
shown in Figure \ref{loopR} for $N=2$. The weight of a non-contractible (resp.\ contractible)
loop on this surface is $\left(S_{h'}/S_{h}\right)^N$ (resp.\ $S_{h'}/S_{h}$), due to the $2\pi N$ winding of non-contractible loops;
this must finally be summed over all possible path in the Dynkin diagram.
For instance, consider the case of Figure \ref{loopR}, with $N=2$ and two non-contractible loops.
The first loop is the boundary between a cluster of height $\hbdy = 1$ on its left and $2$ on its right. It picks up a factor $S_2^2/S_1^2$.
The second loop can either surround a cluster of height $1$ or $3$, and therefore gets a factor $(S_3^2+S_1^2)/S_2^2$.

In the general case, we consider heights living on the $A_m$ Dynkin diagram, and define the following matrix
$\left(\Lambda_N\right)_{i,j}=\delta_{|i-j|,1}\left( S_i / S_j \right)^N$ for $i,j=1,\ldots,m$. Thus, $\Lambda_N$ is
the adjacency matrix with the non-contractible loop weights on $N$ replicas. The matrix element $(\Lambda_N^k)_{\hbdy,h_k}$
is the weight of the configuration with $k$ non-contractible loops, where the boundaries are fixed to $\hbdy$
and the last loop surrounds a cluster of height $h_k$. Since we sum over the height of the last cluster and we
fixed $\hbdy = 1$, the full weight is $\bra{{\bf h}_{\rm free}}\Lambda_N^k\ket{\hbdyvec}$ where
$\bra{{\bf h}_{\rm free}} = \left(1,\ldots,1\right)$ and $\ket{\hbdyvec} = \left(1,0,\ldots,0\right)^T$.
The weight of a set of $k$ contractible loops is then
$w=\sum_{i=1}^m\langle {\bf h}_{\rm free}|\lambda_i\rangle\langle \lambda_i | \hbdyvec \rangle\lambda_i^k$,
where $\ket{\lambda_i}$ and $\bra{\lambda_i}$ are the right and left eigenvectors of $\Lambda_N$ associated
to the eigenvalues $\lambda_i$, for $i=1,\ldots,m$.

We hence need to sum over sectors where the weight of non-contractible loops is given by the different eigenvalues of $\Lambda_N$.
We notice that the characteristic polynomial depends only on the products $\left(\Lambda_N\right)_{i,j}\left(\Lambda_N\right)_{j,i}=1$
(expand by the minors of the first column). The characteristic polynomial is hence unchanged if we replace $\Lambda_N$ by
the usual adjacency matrix, with elements $\Lambda_{i,j}=\delta_{|i-j|,1}$. The spectra of the adjacency matrices of $A_m$ Dynkin
diagrams are $\{\lambda_k =2\cos\frac{k\pi}{m+1}\}_{k=1,\ldots,m}$. The normalized eigenvectors of $\Lambda_N$ are found \cite{BauerSaleur89} as
\small
\bea
\ket{\lambda_k}_i &=& \sqrt{\frac{2}{m+1}}\left(\frac{\sin\frac{ip\pi}{m+1}}{\sin\frac{p\pi}{m+1}}\right)^N \sin \left( \frac{ik\pi}{m+1} \right) \,, \nonumber \\
\bra{\lambda_k}_i &=& \sqrt{\frac{2}{m+1}}\left(\frac{\sin\frac{ip\pi}{m+1}}{\sin\frac{p\pi}{m+1}}\right)^{-N} \sin \left( \frac{ik\pi}{m+1} \right) 
\eea
\normalsize
for $i=1,2,\ldots,m$.
Finally, the RSOS partition function with $N$ replicas and a boundary is a sum of loop partition functions $Z_{N,k}^{\rm loop}$,
where non-contractible loops get a weight $\lambda_k$, i.e., $Z_N^{\rm RSOS}=\sum_{k=1}^{m}\alpha_k Z_{N,k}^{\rm loop}$.
The prefactor $\alpha_k$ can be computed from the eigenvectors of $\Lambda_N$:
\small
\bea \alpha_k&=&\langle {\bf h}_{\rm free}|\lambda_k\rangle\langle \lambda_k| \hbdyvec \rangle\label{detailedalphak} \\
&=& \frac{2}{m+1}\left(\sin\frac{\hbdy k\pi}{m+1}\right)^{1-N}\sum_{i=1}^m\left(\sin\frac{ip\pi}{m+1}\right)^N\sin\frac{ik\pi}{m+1} \,. \nonumber\eea
\normalsize
The dominant contribution comes from non-contractible loop with the largest possible weight, $2\cos{\frac{\pi}{1+m}}$; this is because the corresponding sector is associated with the smallest electric charge. In the limit where the system size goes to infinity we thus have $Z_{N}^{\rm RSOS}\thicksim\alpha_1Z_1^{\rm loop}$.

We note that the detailed coefficient $\alpha_k$ will depend on the boundary condition imposed on the left of the system. For fixed height $\hbdy$, we see that the prefactor in (\ref{detailedalphak}) contributes a term $\ln\big(\sin\frac{\hbdy k\pi}{m+1}\big)$. Recall now the expression
(see e.g.\ \cite{Freden}) of the Affleck-Ludwig entropy \cite{AffLud}---we restrict here to the unitary case $p=1$ for simplicity:
\small
\begin{equation}
g_{1\hbdy}=\left[{2\over m(m+1)}\right]^{1/4}\left[ 2{\sin{\pi\over m}\over\sin{\pi\over m+1}}\right]^{1/2}\sin{\pi \hbdy\over m+1} \nonumber
\end{equation}
\normalsize
We see that the $\hbdy$ dependence of the $O(1)$ contribution to the R\'enyi entropy matches the (logarithm of) the degeneracy factor $g_{1\hbdy}$.
Meanwhile, it is well known that fixing the RSOS height to $\hbdy$  corresponds to the boundary condition $(1\hbdy)$ in the above notation, while it is also known that the conformal boundary condition contributes to the entanglement by a factor $O(1)$ which is precisely the logarithm of the degeneracy factor---the Affleck-Ludwig entropy \cite{AffLud}.
Our calculation thus reproduces this subtle aspect of the entanglement entropy as well. 

We also note that, despite the relative freedom offered by the coefficients $\alpha_k$, there does not seem to be any satisfactory way to concoct a boundary condition for which the leading term $\alpha_{k=1}$ cancels out for all $N$.

\subsection{The non-compact case}

As an example of non-compact CFT  we consider the $c=1$ Liouville theory, which can be obtained by taking the $m\to\infty$ limit of
the unitary CFTs based on the $A_m$ RSOS models \cite{RunkelWatts}. Using our lattice approach, it is easy to see which features
might emerge in this limit. Indeed, going back to the calculation in the preceding subsection, and writing  the contributions from all
possible loop weights, we get the partition function for the $N$-replica model in the form
\begin{equation}
Z_N\propto L^{-{1\over 6}(N-{1\over N}) (1-6e_0^2/g)}\sum_{k=1}^m c_{k,N} L^{-{e_0^2\over g} {k^2-1\over N}} \,,
\label{ncZN}
\end{equation}
and 
\begin{equation}
Z_1\propto \sum_{k=1}^m c_{k,1} L^{-{e_0^2\over g} (k^2-1)} \,.
\label{ncZ1}
\end{equation}
The coefficients $c_{k,N}$ are difficult to evaluate: they depend not only on the combinatorics of the model, but also on the normalization
in the continuum limit of the different insertions  of lattice vertex operators necessary to give the correct weights to non-contractbile loops. 
Recall $e_0={1\over m+1}$ and $g={m\over m+1}$. We now take the limit $m\to \infty$, following the construction of \cite{RunkelWatts}.
To this end, we have to make an ansatz for the coefficients $c_{k,N}$. Many comments in the literature suggest that the dependency on
$N$ is negligible. Assume for extra simplicity that the $c_{k,N}$ are essentially constant as a function of $k$ as well (this is all up to a lattice-cutoff power-law dependency, which we put in the $L$ term). We have then, replacing sums by integrals when $m$ is large, that
\begin{equation}
{Z_N\over Z_1^N}\sim L^{-{1\over 6}(N-{1\over N})} \frac{\int_0^\infty {\rm d}x \, L^{-x^2/N}}{\left(\int_0^\infty {\rm d}x \, L^{-x^2}\right)^N} \,. \label{ncZNint}
\end{equation}
Note that we have extended the integral to infinity, while since obviously $x\propto {k\over m}$, it looks like it should run only up to $x=1$.
There are two reasons for this: one is that at large $L$ the behavior is dominated by the region of $x$ small anyhow. The other is that we
have, in fact, neglected all the contributions occurring from electric charges (in the lattice derivation) shifted by integers. Accepting (\ref{ncZNint}) we
find, after evaluating the Gaussian integrals,  the result
\begin{equation}
 {Z_N\over Z_1^N}\sim L^{-{1\over 6}(N-{1\over N})} (\ln L)^{{N-1\over 2}} \,. \label{ncZNdoneint}
\end{equation}
Note that there are in fact additional factors of $m$ cropping up when we transform the sums (\ref{ncZN})--(\ref{ncZ1}) into integrals.
They will only affect the entanglement by $O(1)$ terms, so we have neglected them.
 
Finally, taking minus the derivative of (\ref{ncZNdoneint}) at $N=1$ to get the EE we obtain
\begin{equation}
 S={1\over 3}\ln L-{1\over 2}\ln(\ln  L) \,,
\end{equation}
whereas for the R\'enyi entropy we get
\begin{equation}
 S^{(N)}={N+1\over 6N}\ln L-{1\over 2}\ln(\ln L) \,.
\end{equation}
 
Note that the argument hinges crucially on the absence of a non-trivial (power-law) dependency of the $c_{k,N}$ on $k$. Since these coefficients depend, in part, on the correspondence between lattice and continuum, this  may well provide a non-universal contribution to the $\ln(\ln L)$ term.

\end{document}